\documentclass[openacc]{rstransa}

\titlehead{Research}

\begin{document}

\title{Spectropolarimetric analysis of waves linked to FIP}

\author{
M. Murabito$^{1}$, M. Stangalini$^{2}$}

\address{$^{1}$INAF - Osservatorio Astronomico di Roma, Via Frascati 33, Monteporzio Catone, Rome, Italy\\
$^{2}$Italian Space Agency, Via del Politecnico, SNC, Rome, Italy}

\subject{Astrophysics}


\corres{Mariarita Murabito\\
\email{mariarita.murabito@inaf.it}}

\begin{abstract}

High-resolution spectropolarimetry has opened new avenues for understanding how chromospheric waves shape coronal plasma composition. All modeling efforts so far highlight wave activity and, in particular, the ponderomotive force associated to Alfvénic perturbations, as a key ingredient. Over recent years, studies based on spectropolarimetric measurements in the solar chromosphere have identified magnetic perturbations associated to waves linked to regions in the corona with enhanced FIP bias. 

{Building on this established framework, the present work explores the diagnostic potential of Stokes V amplitude asymmetries as an alternative diagnostic tool for investigating wave behavior relevant to compositional fractionation processes.
}

\end{abstract}




\maketitle

\section{Introduction}
An intriguing aspect in the physics of the solar atmosphere is the long-standing discrepancy in elemental abundances between the solar corona and the photosphere (e.g., \cite{Widing1989,Sheeley1995,Sheeley1996}). In the corona, elements with a First Ionization Potential (FIP) below 10 eV are typically enriched by factors of 3–4 relative to their photospheric abundances. This phenomenon, known as the FIP effect, is quantified through the FIP bias, defined as the ratio of an element’s atmospheric abundance to its photospheric value. For a discussion on this topic see Brooks et al. 2025 in this topical issue. It was suggested based on theoretical models \cite{Laming2004} that FIP bias could be linked to Alfvénic waves. {Further studies have shown that the FIP bias arises from the ponderomotive force generated by MHD waves as they reflect and refract within the chromosphere \cite{Laming2015,Laming2017}. This force acts preferentially on low-FIP ions, which are present only in the lower atmospheric layers where partial ionization occurs. As a result, it plays a key role in producing the abundance anomalies observed in the fully ionized coronal plasma relative to the photosphere.}\\
Independent work aimed at the detection of magnetic oscillations in the solar atmosphere using high resolution spectropolarimetric measurements turned out to be a possible first confirmation of the above scenario \cite{Stangalini2021}. {Using chromospheric circular polarization measurements, oscillatory signals were detected in the umbra of AR 12546, primarily in the 3-min band. Their frequency dependence, spatial localization, and phase relationship with intensity were shown to be inconsistent with opacity effects or instrumental cross-talk alone, suggesting a genuine magnetic origin of the signal. These 
chromospheric magnetic perturbations} were spatially associated with regions of enhanced FIP bias, offering the first observational confirmation of the mechanisms long predicted by theory \cite{Baker2021}. This work marked the beginning of a series of publications aimed at deepening our understanding of the mechanism underlying this link. Among other things, these works for instance investigated the magnetic field geometry of the regions associated with FIP \cite{Murabito2021} and the wave propagation and reflection, {for first time,} at those regions \cite{Murabito2024}. 

Waves are observed to periodically shift, broaden, strengthen and modify the asymmetry of Stokes-Q and -V profiles \cite{PlonerSolanki1997}. An alternative spectropolarimetric diagnostic that could serve as a practical proxy, comparable to the quantities used in previous studies, is the Stokes-V asymmetry (either amplitude or area). This quantity, derived solely from Stokes-V measurements, can {arise from the combined effect of line-of-sight velocity and magnetic-field gradients \cite{Illing1975,Solanki1993}}. In particular, according to \cite{Solanki1993} and \cite{PlonerSolanki1997}, the sign of the Stokes-V asymmetry (area or amplitude) is determined by the following relation: 

\begin{equation}
\label{eq1}
sign(\delta a) = sign({-\frac{\partial v}{\partial \tau}}{\frac{\partial |B|}{\partial \tau}})
\end{equation}

where the two terms correspond to the gradients of the line-of-sight velocity and magnetic field, respectively. 

In this contribution, we extend our {previous }analysis {of the same dataset} by investigating {whether} Stokes-V amplitude asymmetry {can be used as an alternative proxy} to identify regions in the solar chromosphere {potentially} associated with FIP enhancements in the corona. This diagnostic could provide a practical alternative to previous methods, requiring less demanding observations and avoiding some of the more complex analysis techniques. 


\maketitle

\section{Data set and methods}

The dataset consists of high-resolution, full-Stokes spectropolarimetric measurements of the chromospheric Ca II 854.2 nm line in the sunspot portion of active region (AR) 12546, acquired with IBIS instrument \cite{Cavallini2006} at the Dunn Solar Telescope. The chromospheric line was sampled at 21 wavelength points with spectral steps of 60 m\AA\ and a temporal cadence of 48 s. A full description of the data used can be found in \cite{Stangalini2018} and \cite{Murabito2019}. These observations are available at the IBIS-A data archive \cite{Ermolli2022}.

We calculate the Stokes-V amplitude asymmetry from the Ca II Stokes-V measurements using the following formula: 

\begin{equation}
\delta a = \frac{a_{b} - a_{r}}{a_{b} + a_{r}}
\end{equation}

where a$_{b}$ and a$_{r}$ are the unsigned amplitudes of the blue and red wings of Stokes-V respectively, {defined as the absolute value of the maximum (blue) and minimum (red) of the Stokes-V signal.}

\section{Results}

{We perform time-averaged Stokes-V amplitude asymmetry over the full observational window, which is significantly longer than the typical wave periods, to filter out rapid oscillation associated with short-scale waves. In this regime, the term $\frac{\partial B}{\partial \tau}$ in Eq. \ref{eq1}, typically accounts for projection effects or changes in magnetic polarity along the line-of-sight. Since our observations are taken at disk center, thus projection effects are negligible. Furthermore, previous findings (e.g., \cite{Murabito2021}) have shown that the magnetic field expansion gradient ($\frac{\partial B}{\partial \tau}$) maintains a consistent sign across both sides of the sunspot. Under these conditions, the sign of the averaged Stokes-V amplitude asymmetry directly reveals the persistent velocity gradients ($\frac{\partial v}{\partial \tau}$) associated with stable oscillations.}

\begin{figure}[!h]
\centering

\includegraphics[scale=0.345, clip, trim= 0 50 0 50]{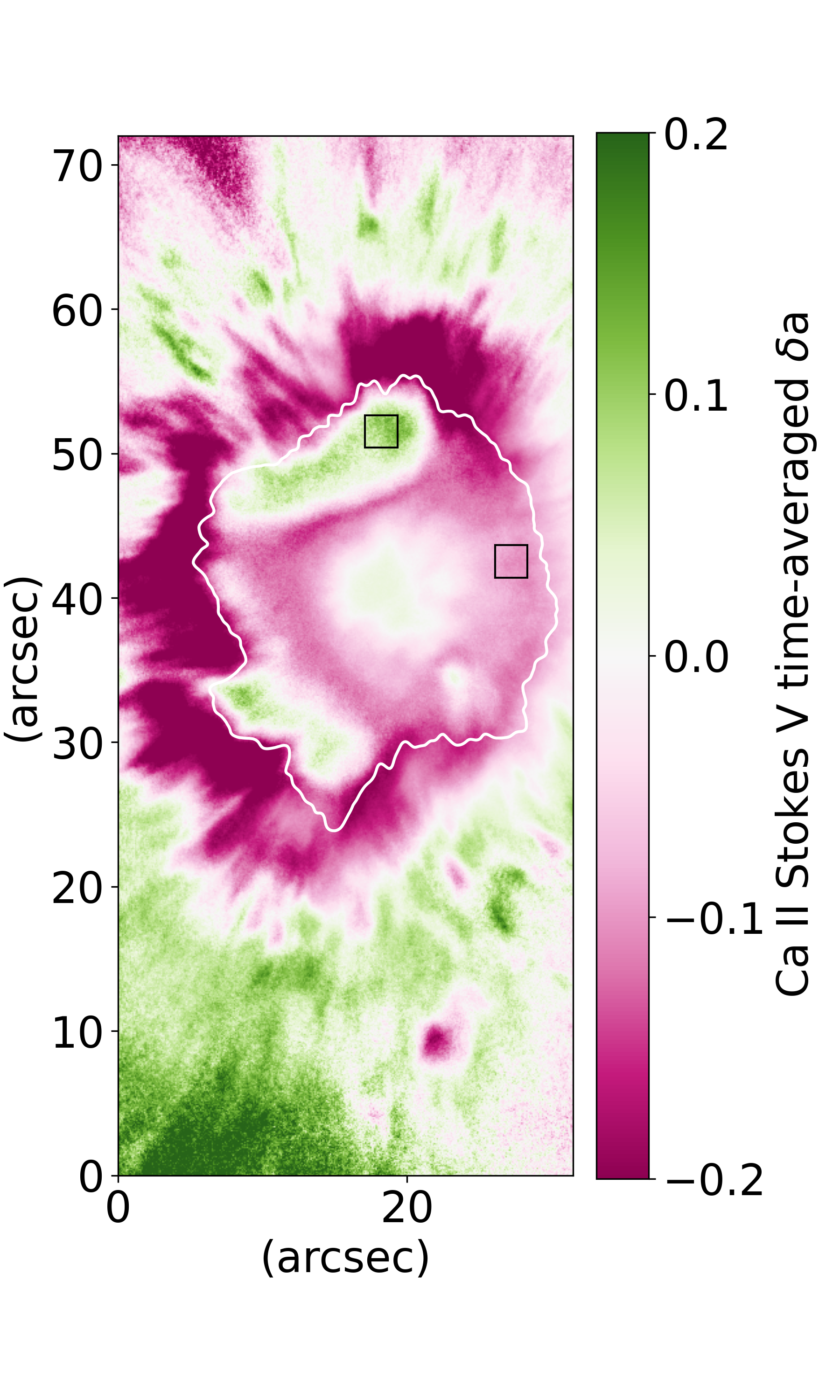}
\includegraphics[scale=0.30, clip, trim= 0 0 0 28]{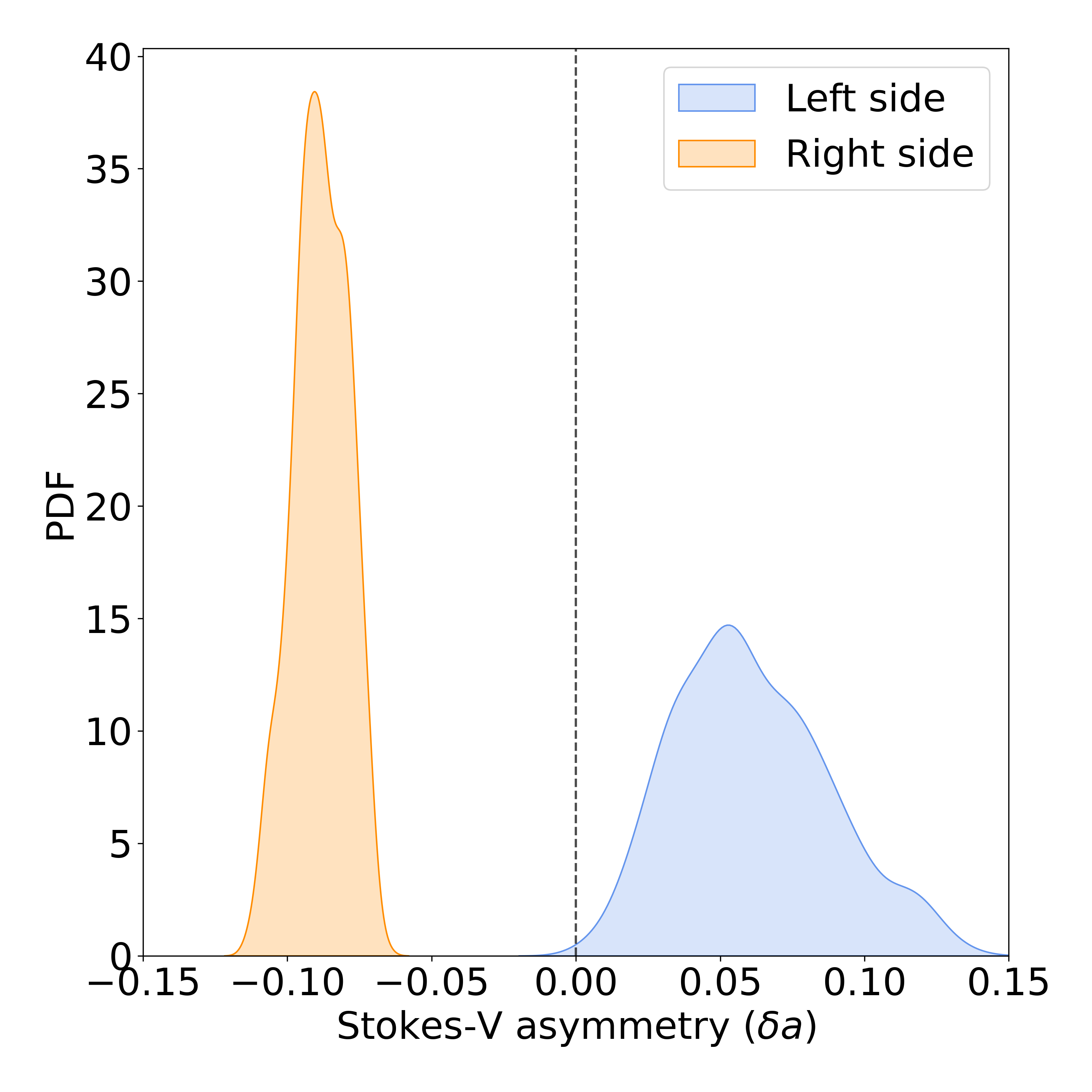}

\caption{Left: Time-averaged Stokes-V $\delta$a map. White contours indicate the umbra–penumbra boundary, derived from the continuum Fe I 617.3 nm intensity line. {Right: PDF of Stokes-V $\delta$ sign in the two boxed shown in the left map.} }
\label{fig1}
\end{figure}

\begin{figure}[!h]
\includegraphics[scale=0.34, clip, trim = 80 0 130 40]{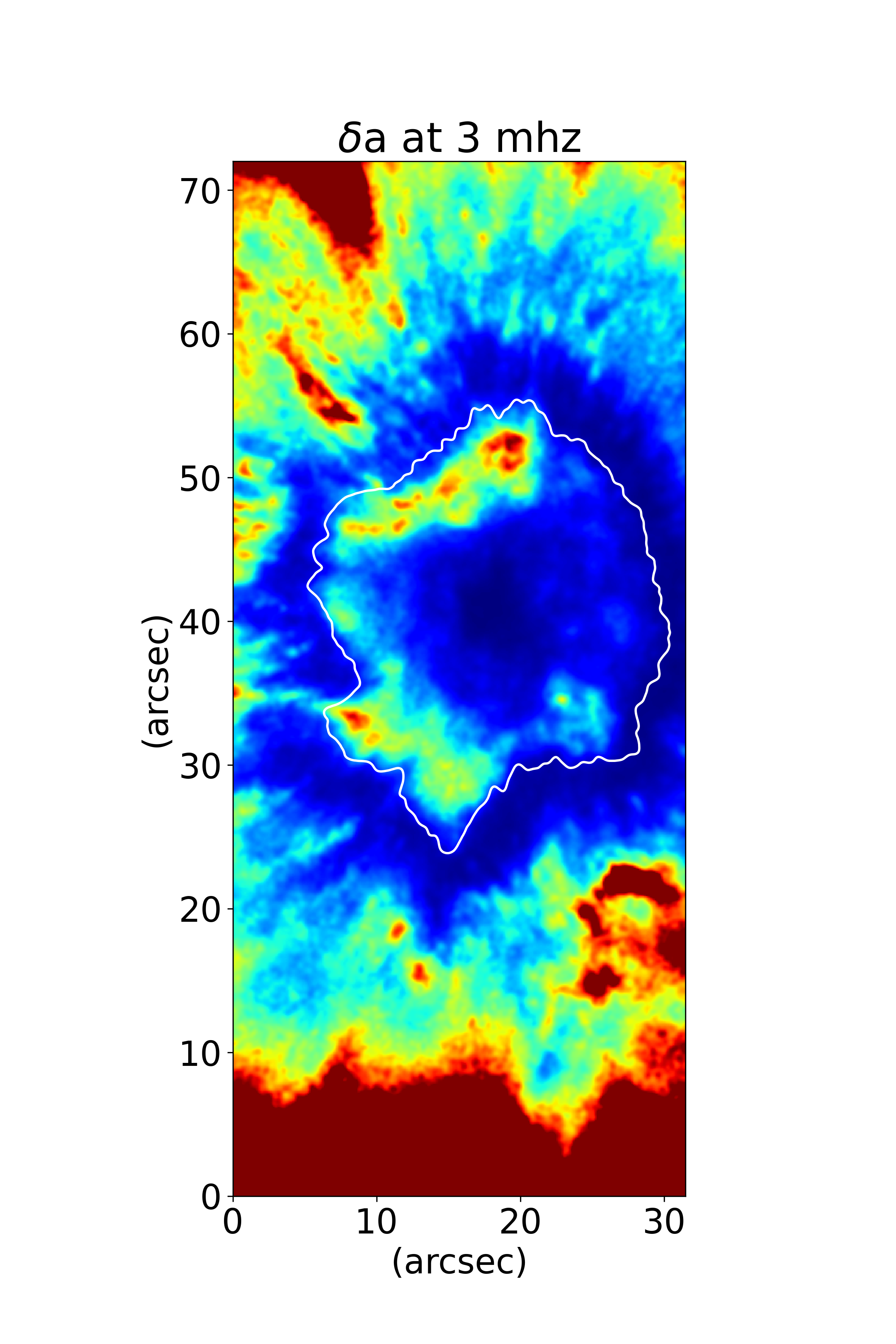}
\includegraphics[scale=0.34, clip, trim = 100 0 130 40]{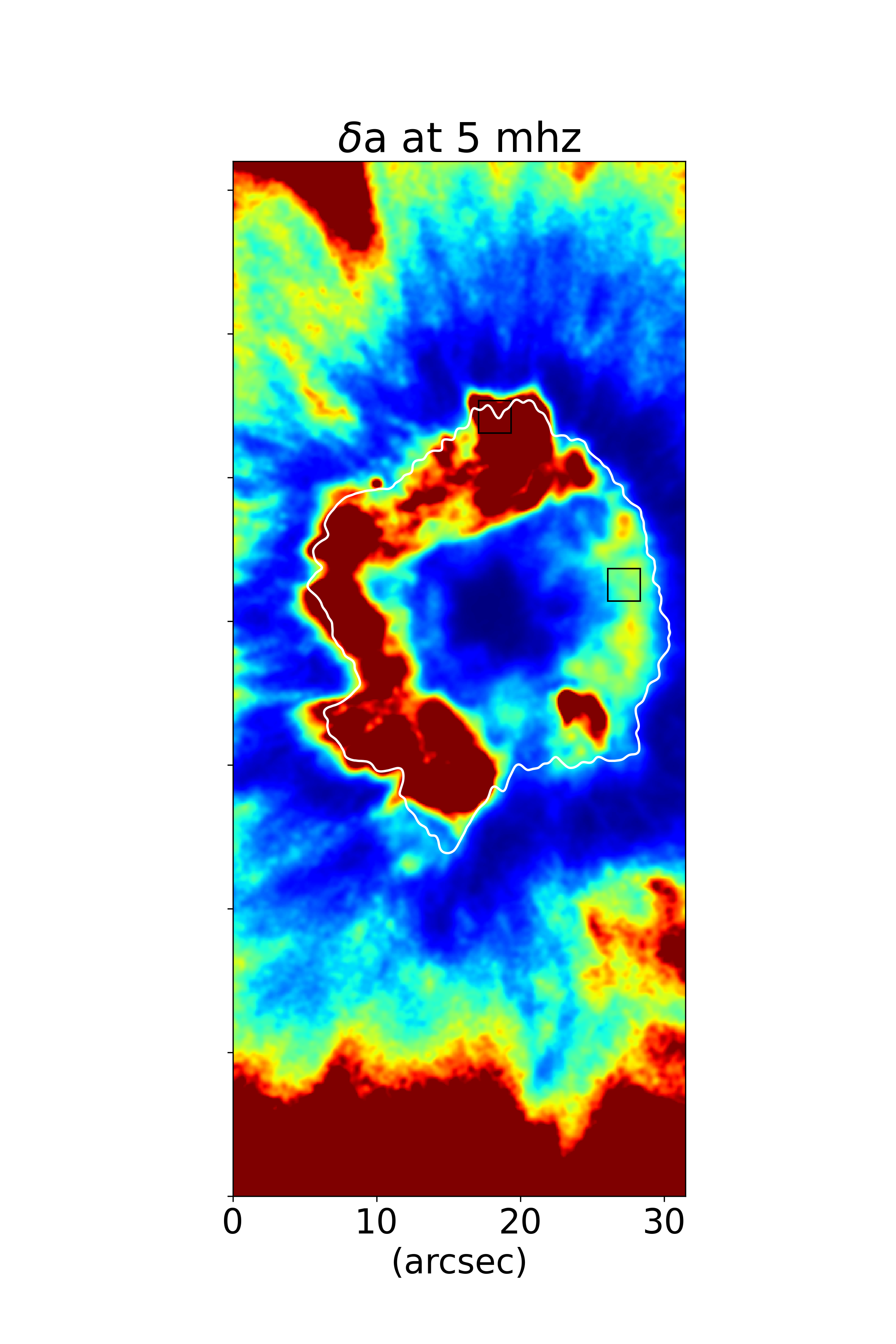}
\includegraphics[scale=0.34, clip, trim = 100 0 130 40]{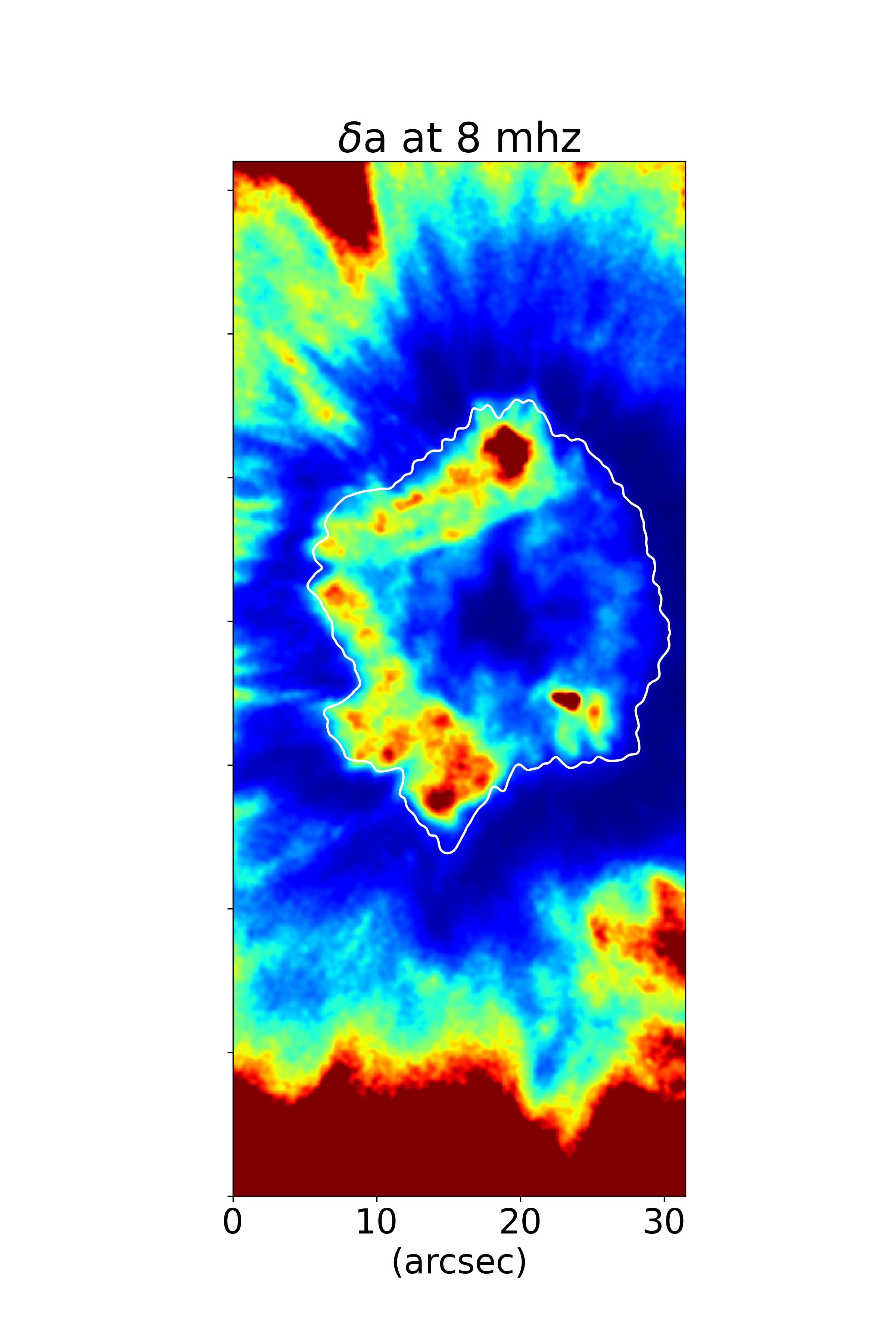}
\caption{Power maps extracted through Fourier analysis of the Stokes-V $\delta$a time series at 3, 5 and 8 mHz, each one averaging over a 1 mHz {bandwidth}. White contours denote the continuum umbra-penumbra boundary. The two black boxes indicate where we compute the averaged power spectrum displayed in the Figure 3.}
\label{fig2}
\end{figure}

{Figure 1, (left panel)} reports the time-averaged Stokes-V amplitude asymmetry map. 
{The map displays a clear asymmetry ring at the umbra-penumbra boundary which is consistently negative, though it appears enhanced (more negative) on the left and upper-central side. Inside the umbra, however, we observe a distinct spatial reversal: the upper-left side of the umbra exhibits a positive asymmetry, whereas the right side shows negative values. This spatial distribution of the ring is consistent with the expansion factor and magnetic field gradient maps reported by \cite{Murabito2021}, who demonstrated trough dedicated photospheric and chromospheric inversions, faster expansion of field lines in locations where magnetic perturbations were observed.}

{To further quantify the behavior inside the umbra, the right panel of Figure 1 shows the PDF of the asymmetry measured within the two regions of interest (boxes in the left Stokes-V asymmetry map). The distinct bimodal behavior of the Stokes-V asymmetry sign, shifting from positive to negative across the umbra, may be ascribed to waves possibly interacting with a reflection layer in the solar atmosphere, a result consistent with findings by \cite{Murabito2024}.}

To investigate the wave power, the Stokes-V amplitude asymmetry was filtered in the Fourier space. Figure 2 shows power maps of the filtered Stokes-V amplitude asymmetry in the 3, 5 and 8 mHz bands (each one averaging over a 1 mHz {bandwidth}). These maps reveal the same peculiar circular pattern inside the umbra detected by Stangalini \cite{Stangalini2021}, with the highest oscillatory power found in the 5-mHz band. In Figure 3 we show the mean spectra obtained by averaging the power spectrum over the two 25$\times$25 pixel boxes indicated in the middle panel of Figure 2. Overall, the mean power spectrum on the left side (blue curve) is enhanced compared with that on the right side (orange curve). We detect an enhancement in the 4–5 mHz band, with amplitude variations between the two regions, which is particularly prominent on the left side of the umbra. We also observe a slight broadening of the 4–6 mHz bands toward lower frequencies on the left side, along with a shift of the multiple peaks on the right side. The multiple oscillatory peaks in the 4–8 mHz band are enhanced by more than a factor of three on the left side, contributing to the stronger signal observed in the power maps in this region of the umbra.

\begin{figure}[!h]
\centering
\includegraphics[width=4in]{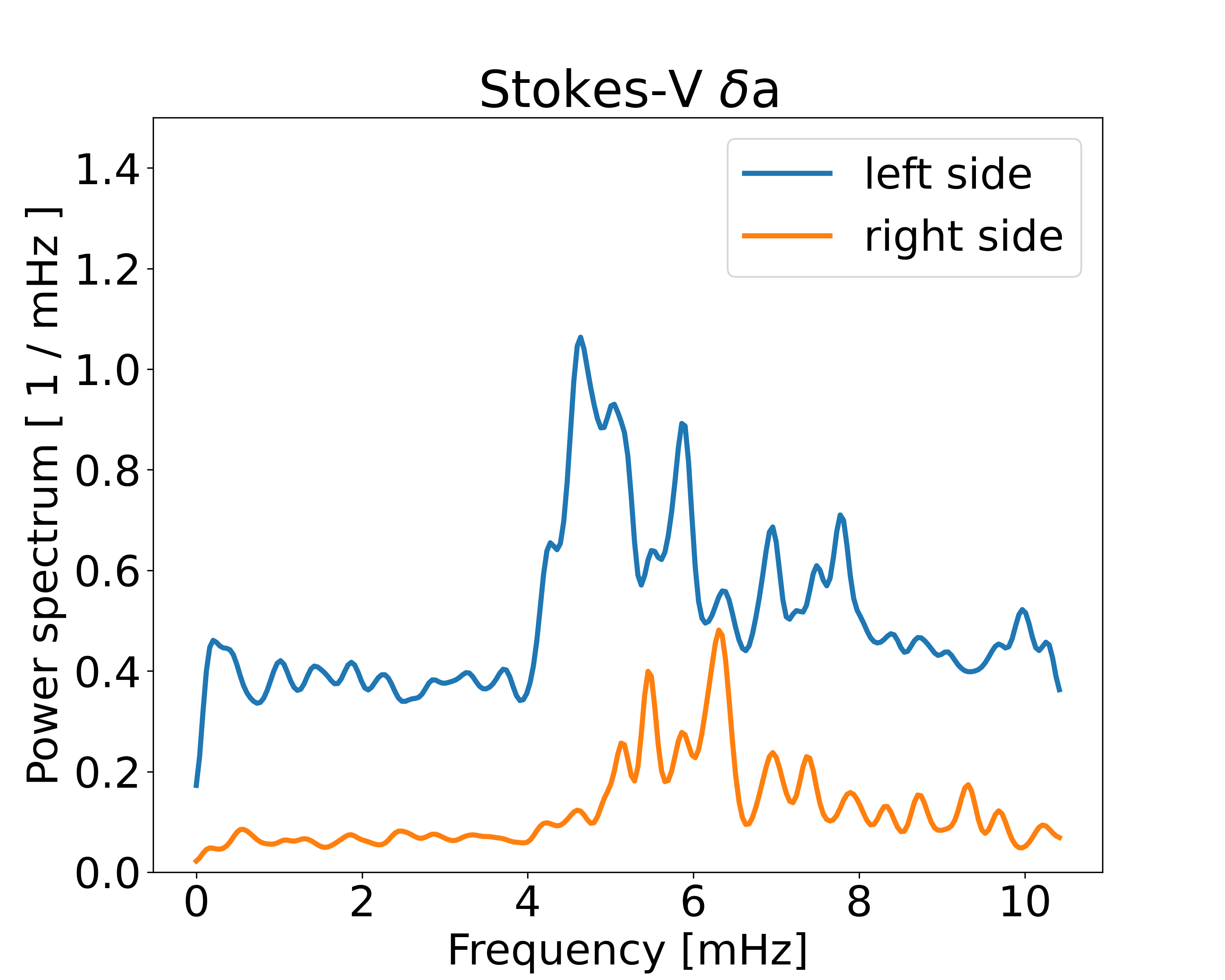}
\caption{Ca II Stokes-V asymmetry power spectrum. The blue and orange curve indicate the average power spectrum inside the two 25$\times$25 pixel boxes shown in the middle panel of Figure 2. }
\label{fig4}
\end{figure}

\section{Discussion and Conclusion}

AR 12546, for its exceptional size, magnetic field strength and long-term stability, provides an ideal benchmark for studying the coupling between coronal composition and waves. Its stability allow us to exclude effects from the intrinsic evolution of the structure and focus solely on the wave dynamics. 
{The magnetic oscillations observed in the umbra of this exceptional data \cite{Stangalini2021}, were associated with regions of strong coronal FIP bias \cite{Baker2021} and were found to be reflected \cite{Murabito2024} as predicted by the Laming's model \cite{Laming2019}}. Unfortunately these remain the only available data suitable for such an analysis, as it was long assumed that the driver of MHD waves reflection and/or refraction originate from the corona. In this context, the study of Mesoraca et al. 2025, in this topical issue, {extends these earlier investigations by introducing a temporal perspective.} Following an AR across the solar disk, they found that oscillations {of photospheric line-of-sight velocity }in the 4-5 mHz band in the umbra appear recurrently and seem to correspond to changes in the FIP-bias distribution.

In this contribution, we further exploit this unique chromospheric spectropolarimetric dataset to investigate the potential of Stokes-V amplitude asymmetry as an alternative proxy. {Oscillations in the 4–6 mHz band are particularly enhanced on the left and upper-central side of the umbra. These regions coincide with areas where magnetic oscillations associated with FIP enhancement have been previously reported. The time-average asymmetry highlights different sign between the left and right side of the umbra, which may be attributed to a reflection layer in the solar atmosphere. All together, these findings are in agreement with the results reported by \cite{Stangalini2021,Murabito2024}.} 

{This alternative approach facilitates studies of waves–composition coupling by relaxing the observational requirements typically associated with dedicated wave diagnostics. In particular, it reduces the need for long, high-cadence time series with very high polarimetric sensitivity. Moreover, it reduces the dependence on diagnostic techniques that rely on specific observing configurations and stringent signal-to-noise requirements to prevent measurements from being dominated by noise, a limitation affecting several spectropolarimetric  diagnostics, including inversions. The Stokes-V amplitude asymmetry can be extracted from datasets with a limited number of spectral points and shorter time coverage, making it a practical diagnostic in a wider range of observational contexts.}

These results, {as in previous studies based on the same dataset, emphasize the importance of chromospheric spectropolarimetric measurements for understanding the nature and variability of elemental abundance between the corona and the photosphere.} 
However, the IBIS and EIS datasets used here were only serendipitously acquired and almost simultaneously. Coordinating ground-based spectorpolarimetric 
chromospheric measurements with space-based coronal observations remains a major challenge. The approach described in these results, developed over several years of collaboration, requires resolving and measuring the magnetic field strength and topology at the smallest scales in a layer not yet fully modeled or understood. Continuous, high resolution, high cadence spectropolarimetric coverage from space is currently limited at photospheric layers. The extremely limited number of simultaneous ground-based and space observations demonstrates the difficulty in obtaining useful datasets suitable for studying the link between Alfvénic waves and FIP.\\
Finally, the recent work by Berretti et al. \cite{Berretti2025} identified unusual 5 mHz oscillation in a small portion of a large statistical sample of photospheric ARs. This suggests that the wave dynamics within a single AR can change over time, resulting in the intermittent appearance of frequencies associated with MHD waves. This raises the question of whether the ponderomotive force at the base of the FIP bias is an intermittent feature of ARs, leading to the intermittent appearance of FIP in the higher layers of the solar atmosphere. We believe that understanding this aspect is crucial for comprehending the mechanisms underlying the FIP bias itself and could be addressed in the future by studying in greater detail the evolution of wave dynamics over time and its relationship to FIP, although based on photospheric data.

\ack{The authors wish to acknowledge scientific discussions with the Waves in the Lower Solar Atmosphere (WaLSA; www.WaLSA.team) team. This research has received funding from the European Union’s Horizon 2020 Research and Innovation program under grant agreement No 824135 (SOLARNET), and No 739500 (PRE-EST). }



\begin{thebibliography}{99}

\bibitem{Widing1989} 
Widing, K.~G. \& Feldman, U.\ 1989  Abundance Variations in the Outer Solar Atmosphere Observed in SKYLAB Spectroheliograms. \textit{Astrophys. J.}, \textbf{344}, 1046. (doi:10.1086/167871)

\bibitem{Sheeley1995} Sheeley, N.~R.\ 1995  A Volcanic Origin for High-FIP Material in the Solar Atmosphere. \textit{Astrophys. J.}, \textbf{440}, 884. (doi:10.1086/175326)

\bibitem{Sheeley1996} Sheeley, N.~R.\ 1996  Elemental Abundance Variations in the Solar Atmosphere. \textit{Astrophys. J.}, \textbf{469}, 423. (doi:10.1086/177792)

\bibitem{Laming2004} Laming, J.~M.\ 2004  A Unified Picture of the First Ionization Potential and Inverse First Ionization Potential Effects. \textit{Astrophys. J.}, \textbf{614}, 2, 1063. (doi:10.1086/423780)

\bibitem{Laming2015}
Laming JM. 2015 The FIP and Inverse FIP Effects in Solar and Stellar Coronae. \textit{Living Rev. Solar Phys.} \textbf{12}, 2. (doi:10.1007/lrsp-2015-2)

\bibitem{Laming2017}
Laming JM. 2017 The First Ionization Potential Effect from the Ponderomotive Force: On the Polarization and Coronal Origin of Alfvén Waves. \textit{Astrophys. J.} \textbf{844}, 153. (doi:10.3847/1538-4357/aa7cf1)

\bibitem{Stangalini2021} Stangalini, M., Baker, D., Valori, G., \textit{et al.} 2021  Spectropolarimetric fluctuations in a sunspot chromosphere. \textit{Philosophical Transactions of the Royal Society of London Series A}, \textbf{379}, 2190, 20200216. (doi:10.1098/rsta.2020.0216)

\bibitem{Baker2021} Baker, D., Stangalini, M., Valori, G., et al.\ 2021  Alfvénic Perturbations in a Sunspot Chromosphere Linked to Fractionated Plasma in the Corona , \textit{Astrophys. J.}, \textbf{907}, 1, 16. (doi:10.3847/1538-4357/abcafd)

\bibitem{Murabito2021}
Murabito M, Stangalini M, Baker D, Valori G, Jess DB, Jafarzadeh S, Brooks DH, Ermolli I, Giorgi F, Grant SDT, Long DM, van Driel-Gesztelyi L. 2021 Investigating the origin of magnetic perturbations associated with the FIP Effect. \textit{Astron. Astrophys.}, \textbf{656}, A87. (doi:10.1051/0004-6361/202141504)

\bibitem{Murabito2024} Murabito, M., Stangalini, M., Laming, J.~M., \textit{et al.} 2024 Observation of Alfvén Wave Reflection in the Solar Chromosphere: Ponderomotive Force and First Ionization Potential Effect. \textit{Physical Review Letters}, \textbf{132}, 21, 215201. (doi:10.1103/PhysRevLett.132.215201)

\bibitem{Illing1975} Illing, R.~M.~E., Landman, D.~A., \& Mickey, D.~L.\ 1975  Broad-band circular polarization of sunspots: spectral dependence and theory. \textit{Astron. Astrophys.}, \textbf{41}, 2, 183. 

\bibitem{Solanki1993} Solanki, S.~K.\ 1993  Smallscale Solar Magnetic Fields - an Overview. \textit{Space Science Reviews}, \textbf{63}, 1-2, 1. (doi:10.1007/BF00749277)

\bibitem{PlonerSolanki1997} Ploner, S.~R.~O. \& Solanki, S.~K.\ 1997 Influence of kink waves in solar magnetic flux tubes on spectral lines.  \textit{Astron. Astrophys.}, \textbf{325}, 1199.
\bibitem{Houston2020} Houston, S.~J., Jess, D.~B., Keppens, R., \textit{et al.} 2020 Magnetohydrodynamic Nonlinearities in Sunspot Atmospheres: Chromospheric Detections of Intermediate Shocks . \textit{Astrophys. J.}, \textbf{892}, 1, 49. (doi:10.3847/1538-4357/ab7a90)

\bibitem{Cavallini2006} Cavallini, F.\ 2006 BIS: a new post-focus instrument for solar imaging spectroscopy. \textit{Solar Phys.}, \textbf{236}, 2, 415. (doi:10.1007/s11207-006-0103-8)

\bibitem{Stangalini2018} Stangalini, M., Jafarzadeh, S., Ermolli, I., \textit{et al.} 2018 Propagating spectropolarimetric disturbances in a large sunspot. \textit{Astrophys. J.}, \textbf{869}, 2, 110. (doi:10.3847/1538-4357/aaec7b)

\bibitem{Murabito2019} Murabito, M., Ermolli, I., Giorgi, F., \textit{et al.} 2019 Height dependence of the penumbral fine-scale structure in the inner
solar atmosphere. \textit{Astrophys. J.}, \textbf{873}, 2, 126. (doi:10.3847/1538-4357/aaf727)

\bibitem{Ermolli2022} Ermolli, I., Giorgi, F., Murabito, M., et al.\ 2022, \textit{Astrophys. J.}, \textbf{661}, A74. (doi:10.1051/0004-6361/202142973)

\bibitem{Laming2019} Laming, J.~M., Vourlidas, A., Korendyke, C., \textit{et al.} 2019 Element Abundances: A New Diagnostic for the Solar Wind. \textit{Astrophys. J.}, \textbf{879}, 2, 124. (doi:10.3847/1538-4357/ab23f1)

\bibitem{Berretti2025} Berretti, M., Stangalini, M., Verth, G., et al.\ 2025  Umbral oscillations in the photosphere: A comprehensive statistical study.  \textit{Astron. Astrophys.}, \textbf{697}, A156. (doi:10.1051/0004-6361/202453176)



\end{thebibliography}
\end{document}